\documentclass[preprint,12pt,authoryear]{elsarticle}
 \usepackage{amssymb}
 \usepackage{natbib}
\usepackage{amsfonts,amsmath,epsfig,psfrag}
 \usepackage{amsmath}
 \usepackage{tikz}
 \usepackage{float}
\usepackage{pgflibraryshapes}
 \usepackage{graphicx,psfrag}
\usepackage[normalem]{ulem}
 \makeatletter \oddsidemargin 0.0cm \evensidemargin 0.0cm
\newcommand{\be}{\begin{equation}}
\newcommand{\en}{\end{equation}}
\newcommand{\la}{\label}

\newcommand{\paa}{\partial}
\def\rr#1{(\ref{#1})}
\def\bm#1{\mbox{\boldmath{$#1$}}}

\def\ii{{\rm i}}
\def\ee{\mathrm{e}}

\def\r{{ R}}
\def\t{{ T}}
\def\o{{\bf 0}}
\def\q{{ Q}}

\newcommand{\bs}[1]{{\Large\textsf{\textbf{#1}}}}

\begin{document}

\begin{frontmatter}
\title{\bs{A refined model for the buckling of film/substrate bilayers}}
\author[mymainaddress]{Guan Wang }
\author[mymainaddress]{Yang Liu }
\author[secondaryaddress]{Yibin Fu\corref{mycorrespondingauthor}}
\cortext[mycorrespondingauthor]{Corresponding author: y.fu@keele.ac.uk}
\address[mymainaddress]{Department of Mechanics, Tianjin University, Tianjin 300072, China}
\address[secondaryaddress]{School of Computing and Mathematics, Keele University, Staffs ST5 5BG, UK}

\begin{abstract}
The classical reduced model for a film/substrate bilayer is one in which the film is governed by the Euler-Bernoulli beam equation and the substrate is replaced by an array of springs (the so-called Winkler foundation assumption). We derive a refined model in which the normal and shear tractions at the bottom of the film are expressed in terms of the corresponding horizontal and vertical displacements, and the response of the half-space is described by the exact theory. The self-consistency of the refined model is confirmed by showing that it yields a four-term (in the incompressible case) or
six-term (in the compressible case) expansion for the critical strain that agrees with the expansion given by the exact theory. %
%
%
%
%
%
%
\end{abstract}
\end{frontmatter}

\section{Introduction}
The buckling and post-buckling of a film/substrate bilayer, which is often idealised as a coated hyperelastic half space, has received much attention in recent decades due to applications
ranging from cell patterning \citep{dw2020}, optical gratings \citep{lee2010, ma2013, kim2013}, and creation of surfaces with desired wetting and adhesion properties \citep{chan2008, yang2010, zhang-gao2012}, to the deduction of material properties of ultrathin films \citep{stafford2004, chan2009}. Under the framework of nonlinear elasticity, early studies include the linear analyses  by \citet{dn1980}, \citet{sks1994}, \citet{os1996}, \citet{bon1997}, \citet{so1997},  and the nonlinear post-buckling analysis by \citet{cai-fu1999}. More recent studies have addressed the phenomenon of period doubling \citep{fu-cai2015}, effects of pre-stretching in the half-space \citep{hutchinson2013}, compressibility \citep{ld2014,cai-fu2019}, multi-layering \citep{czh2014,wzn2020,zcf2022}, growth \citep{AKG2019,AFG2020,lzdc2020}, and magnetorheology \citep{rd2021}. There also exists a large body of literature based on various approximate models for the layer and substrate; see, e.g., \citet{CH2004}, \citet{HH2005}, \citet{AB2008}, \citet{zch2015}, and the references therein.

With the use of the exact theory of nonlinear elasticity,   \citet{cai-fu1999} derived the following asymptotic expansion for the critical strain when a neo-Hookean film bonded to a much softer neo-Hookean substrate is subjected to a uni-axial compression:
\be
1-\lambda= \frac{r}{2 k h}+\frac{1}{12} (k h)^2-\frac{13}{480} (k h)^4-\frac{3}{8} \cdot \frac{ r^2}{ (kh)^2}+O((kh)^5), \la{0.1} \en
where $\lambda$ is the stretch ratio in the direction of compression (so that $1-\lambda$ is the strain), $k$ is the wave number of the buckling mode, $h$ is the layer thickness, and $r=\mu_{\rm s}/\mu_{\rm f}$ with $\mu_{\rm f}$ and $\mu_{\rm s}$ denoting the shear moduli for the film and substrate, respectively. Note that the above expansion is only valid when $r =O((kh)^3)$, in which case the first two terms on the right hand side of \rr{0.1} are of the same order and the last two terms are of the same order. Solving this expansion together with $d \lambda/d (kh)=0$, we may determine the critical stretch and associated wave number where the strain $1-\lambda$ attains its minimum.
This two-term result extends the leading order result of  \citet{allen1969} for the plane-stress case.

It can be shown that the leading order version of \rr{0.1}, with only the first two terms on the right hand side retained, can be derived from the classical model
\be
\frac{EI}{1-\nu^2} y''''(x)+\frac{Eh}{1-\nu^2} (1-\lambda) y''(x)=-\gamma k y, \;\;\;\; \gamma=\frac{2 E_s (1-\nu_s)}{(1+\nu_s)(3-4 \nu_s)}, \la{classical} \en
by taking the incompressibility limit $\nu, \nu_s \to 1/2$,
where $E$ and $\nu$ are the Young's modulus and Poisson's ratio for the layer, respectively,  $E_s$ and $\nu_s$ their counterparts for the half-space.
It was shown in \citet{cai-fu2000} that equation \rr{classical} can be derived from the exact theory of nonlinear elasticity using a self-consistent asymptotic procedure; see also discussions in
\citet{AB2008}. When the half-space is absent, the right hand side of \rr{classical} is zero, and the resulting equation is the well-known Euler-Bernoulli beam equation, which has been justified in many studies; see, e.g., \citet{so2014}, \citet{destrade2016}, \citet{steigmann2007}, and the references therein.

For a stand-alone beam that is only subject to a force $P$ along its axis, the Timoshenko beam theory is known to provide a self-consistent refinement on the Euler-Bernoulli beam theory \citep{ti1921, elishakoff2015}.
It is then tempting to model the response of the layer using the Timoshenko beam theory but still model the response of the half-space using the Winkler assumption. This possibility has indeed been analysed by \citet{erbas2021} for the dynamic case without prestress; see also \citet{kps2019}. Adapting equation (17) in \citet{gkn1993} to our static case, we obtain
 \be \frac{E I}{1-\nu^2} \left\{ y''''+  \frac{(8-3 \nu) }{40 (1-\nu)}\, h^2 y^{(6)} \right\}+\frac{E h}{1-\nu^2} (1-\lambda) y''+\gamma k y=0, \la{gkn} \en
 where $y^{(6)}$ denotes the sixth-order derivative of $y$.
However, on substituting a periodic solution $y(x)={\rm e}^{\ii k x}$ into this equation, it is found that the resulting expression for $1-\lambda$ agrees with the first three terms in \rr{0.1}, but not the fourth term. This means that this reduced model has not taken care of the response of the foundation in a self-consistent manner. An earlier attempt at refining the leading order model \rr{classical} was made by \citet{sks1994}. However, as was shown by \citet{cai-fu2000}, their model is not self-consistent either.
The main purpose of the current study is to forsake the Winkler foundation assumption and allow the interface to transmit both vertical and shear tractions so that the resulting refined model can correctly predict the entire right hand side of \rr{0.3}. To this end, we derive a Timoshenko-type plate theory for a prestressed plate using two different approaches, a standard expansion method as used by \cite{cai-fu2000} and the expansion scheme developed in \cite{ds2014} where a consistent plate theory was proposed for compressible hyperelastic materials.  The latter methodology has been applied to derive consistent plate models for incompressible materials \citep{wsd2016}, growing solids \citep{wswd2018}, and nematic liquid crystal elastomers \citep{lmd2020}. Recently, \cite{wsd2019} have shown that the consistent plate model in \cite{ds2014} can recover most existing plate equations by assuming appropriate order relations between the applied load and the aspect ratio. In deriving the above-mentioned Timoshenko-type plate theory for a pre-stressed plate, we naturally obtain an additional connection between the shear traction and the two displacement components at the interface, which then enables us to apply traction and displacement continuity exactly. Thus, our reduced model consists of two differential equations for the displacement components at the interface.

Our reduced model will be derived for a general hyperelastic material, but to illustrate our results, we shall consider the case in which both the film and substrate are described by the strain energy function
\be
W=\frac{1}{2} \mu (I_1-2-2 \log J)+\frac{\mu \nu}{1-2 \nu} (J-1)^2, \la{0.2} \en
where $\mu$ and $\nu$ are the ground state shear modulus and Poisson's ratio, respectively, $I_1$ is the first principal invariant of the Cauchy-Green strain tensors and $J$ expresses the determinant of the deformation gradient. We denote the shear moduli for the film and substrate by $\mu$ and $\mu_{\rm s}$, respectively, but assume that $\nu$ takes the same value for both the film and substrate. A calculation with the aid of {\it Mathematica} \citep{wolfram}, extending the analysis in \citet{cai-fu1999}, reveals that the compressive strain now has the asymptotic expansion
\be
1-\lambda=\frac{2 (\nu-1)^2}{3-4 \nu} \frac{r}{k h} +\frac{1}{12} (kh)^2+d_0 r+d_1 (k h)^4+d_2 r (k h)+d_3 \frac{ r^2}{ (kh)^2}+O((kh)^5), \la{0.3} \en
where
\be d_0=\frac{(1-\nu ) (1-2 \nu)}{3-4
   \nu},\;\;\;\; d_1=\frac{31 \nu ^2+33
   \nu - 40 \nu ^3-29 }{1440 (\nu -1)^2},\la{0.4} \en
   \be d_2=\frac{23-104
   \nu +156 \nu^2 -80 \nu^3}{24 (3-4 \nu )^2},\;\;\;\; d_3=\frac{2 (\nu -1)^2 \left(32
   \nu ^4-60 \nu ^3+47 \nu
   ^2-21 \nu +5\right)}{(4 \nu
   -3)^3}. \la{0.5} \en
As expected, the expansion \rr{0.3} recovers \rr{0.1} in the incompressible limit $\nu \to 1/2$. Our aim is to derive a refined model that will yield the first six terms in \rr{0.3} exactly. We shall also demonstrate that including the next term in the asymptotic expansion \rr{0.1} or \rr{0.3} can significantly improve its accuracy when the materials are incompressible or nearly incompressible.

We note that the term $d_0 r$ in the expansion \rr{0.3} is of order $(kh)^3$, which has no counterpart in the incompressible limit.
As a quick comparison, we note that the model adopted by  \citet{sks1994} would give the following expansion:
\be
1-\lambda=\frac{2 (\nu-1)^2}{3-4 \nu} \frac{r}{k h}+\frac{1}{12} (kh)^2+d_0 r +\frac{  (\nu-1)^2}{2 (3-4 \nu)} r (kh)+O((kh)^6). \la{0.33} \en
This expression correctly predicts the term $-d_0 r$ but differs from \rr{0.3} in the $O((kh)^4)$ terms.

%
%
%

The rest of this paper is organised into four sections as follows. After formulating our problem and deriving the exact bifurcation condition in the next section, we devote Section 3 to the derivation of the above-mentioned refined model. We first derive a Timoshenko-like model for the film and then use Stroh formulation to describe the response of the substrate. In Section 4, we specialise the model to the buckling analysis of a film/substrate bilayer and show that the reduced model can indeed predict \rr{0.3}. The paper is concluded in Section 5 with a summary and a discussion of other possible applications of the current reduced model.
\section{Formulation and the exact bifurcation condition}
\setcounter{equation}{0}
We consider a hyperelastic layer bonded to a hyperelastic half-space. To simplify derivations, we assume for the moment that both the layer and half-space are
{\it compressible} and are in a state of plane strain. The corresponding results for the incompressible case will be obtained by taking an appropriate limit. We first summarise the governing equations valid for both the layer and half-space.

Consider a general, homogeneous elastic body $B$ composed of a
non-heat-conducting elastic material. Three configurations are involved in our analysis:
an initial
unstressed configuration $B_{0}$, a finitely deformed configuration $ B_{e}$, and the current configuration $ B_{t}$ that is obtained by superimposing a small incremental displacement on $B_{e}$.
The position vectors of a representative particle are denoted by ${\bm X}$, ${\bm x}$ and $\tilde{\bm x}$ in $B_{0},
B_{e}$ and $B_{t}$ respectively, and their Cartesian coordinates are denoted by
$X_{A}, x_{i}$ and $\tilde{x}_{i}$.  We write
\be
\tilde{\bm x} ={\bm x} + {\bm u},
\la{2.1}
\en
where ${\bm u}$, with components $u_{i}(x_1, x_2, x_3)$, is the incremental displacement
associated with the deformation $B_{e} \rightarrow B_{t}$.

The deformation gradients arising from the deformations $B_{0}
\rightarrow B_{t}$ and $B_{0} \rightarrow B_{e}$ are denoted by
$\tilde{F}$ and $\bar{\mathbf{F}}$ respectively and defined by
$d\tilde{\bm x}=\tilde{F} d {\bm X}$ and $d {\bm x}=\bar{F} d {\bm X}$.
It then follows that
\be
\tilde{F}_{iA} = (\delta_{ij} + u_{i,j}) \bar{F}_{jA},
\la{2.3}
\en
where here and henceforth a comma indicates differentiation with respect to
the implied spatial coordinate.

It is well-known (see, e.g. \cite{spencer1970}) that in the absence of body forces the incremental equilibrium equations may be written in the form
\be
\chi_{ij,j}=0,
\la{2.9}
\en
where $\chi_{ij}$ are the Cartesian components of the incremental stress tensor and its linearized formula is given by
\be
\chi_{ij}={\cal A}_{jil k} u_{k,l}.
\la{2.16}
\en
The ${\cal A}_{jil k}$ are the first-order
{\it instantaneous elastic moduli}
defined by \citep{co1971}
\be
{\cal A}_{jil k} =  \bar{J}^{-1}\bar{F}_{jA} \bar{F}_{l B} \left.
\frac{\partial^{2} W}{\partial F_{iA} \partial F_{kB}}
 \right|_{F =
{\bar{F}}},
\la{2.12}
\en
where $W$ is the strain-energy function per unit volume. For the isotropic solids under consideration, $W$ can be assumed to be a function of the three principal stretches $\lambda_1, \lambda_2, \lambda_3$. Then the moduli can be computed according to the following formulae \citep{ogden1984}:
\begin{eqnarray}
{\cal A}_{iijj}&=&\lambda_i\lambda_jW_{ij}, \;\;\;\; \hbox{no summation on}\; i \; \hbox{or}\; j, \nonumber \\
{\cal A}_{ijkl}&=&\frac{\lambda_i^2}{\lambda_i^2-\lambda_j^2}
(\lambda_iW_i-\lambda_jW_j)\delta_{ik}\delta_{jl} \nonumber  +\frac{\lambda_i\lambda_j}{\lambda_i^2-\lambda_j^2}
(\lambda_jW_i-\lambda_iW_j)\delta_{il}\delta_{jk}, \;\;\;\;\;
i\ne j,\nonumber
\end{eqnarray}
where $W_i=\paa W/\paa\lambda_i,\,
W_{ij}=\paa^2W/\paa\lambda_i\paa\lambda_j$ etc, and the principal stretches are understood to correspond to the
primary deformation $B_0 \to B_e$. It is straightforward to verify that
\be {\cal A}_{1212}-{\cal A}_{1221}=\bar{\sigma}_1,\;\;\;\; {\cal A}_{2121}-{\cal A}_{2112}=\bar{\sigma}_2, \la{connect} \en
where $\bar{\sigma}_1$ ($=\lambda_1 W_1$) and $\bar{\sigma}_2$ ($=\lambda_2 W_2$) are the principal Cauchy stresses.

In terms of the displacement components and in the absence of body forces, the equilibrium equation then takes the form
\be
{\cal A}_{jil k} u_{k,lj}=0.
\la{equi}
\en
As indicated earlier, the finite deformation associated with
$B_0 \to B_e$ only appears through the moduli defined by \rr{2.12}. Thus, as it stands, the constitutive equation \rr{2.16} is of the same form as that for an anisotropic elastic solid \citep{spencer1970}, and so our results may be adapted to describe anisotropic materials.

For {\it plane strain} problems, $u_3$ is zero and $u_1$ and $u_2$ are only functions of $x_1$ and $x_2$.  
The finite deformations that we consider (e.g. uni-axial compression) are always such that the elastic moduli ${\cal A}_{jil k}$ is zero whenever there is an odd number of 1, 2 or 3 in the suffices. For instance,
$ {\cal A}_{1112}=0, \; {\cal A}_{1123}=0$ etc. As a result, the equilibrium equations in \rr{equi} written out for $i=1, 2$ take the following form:
\be
{\cal A}_{111 1} u_{1,11}+({\cal A}_{112 2}+{\cal A}_{211 2}) u_{2,12}+{\cal A}_{212 1} u_{1,22}=0, \la{equi1} \en
\be
{\cal A}_{121 2} u_{2,11}+({\cal A}_{122 1}+{\cal A}_{221 1}) u_{1,12}+{\cal A}_{222 2} u_{2,22}=0. \la{equi2} \en
The traction vector ${\bm t}$ on any surface with normal pointing in the positive $x_2$-direction has components given by $t_i=\chi_{i2}={\cal A}_{2ilk} u_{k,l}$. Thus, we have
\be t_1={\cal A}_{2112} u_{2,1}+{\cal A}_{2121} u_{1,2}, \;\;\;\;t_2={\cal A}_{2211} u_{1,1}+{\cal A}_{2222} u_{2,2}. \la{tract} \en
We now specialize the above equations to a coated half-space. In the undeformed configurations $B_0$, the coating and half-space are defined by $0 \le X_2 < H, H \le X_2 < \infty$, respectively. We assume that the primary deformation $B_0 \to B_e$ corresponds to a uni-axial compression with stretch $\lambda_1=\lambda$ in the $x_1$-direction. We take $\lambda$ to be our loading/bifurcation parameter. The stretch $\lambda_2$ in the $x_2$-direction can be determined by solving $\bar{\sigma}_2=0$. It follows from \rr{connect}$_2$ that the identity ${\cal A}_{2121}={\cal A}_{2112}$ holds for all values of $\lambda$ and will be used to eliminate the component ${\cal A}_{2121}$ in the subsequent analysis. In the finitely deformed configuration $B_e$, the coating and half-space are defined by
$$0 \le x_2 \le h, \;\;\;\;{\rm and}\;\;h \le x_2 < \infty, $$
respectively, where $h=\lambda_2 H$ is the thickness of the coating layer in $B_e$.

The linearized buckling problem consists of solving the equilibrium equations \rr{equi1} and \rr{equi2} subject to (i) the traction-free boundary conditions ${\bm t} =0$ at $x_2=0$, (ii) continuity of ${\bm t}$ and ${\bm u}$ at the interface $x_2=h$, and (iii) the decaying condition ${\bm u} \to {\bm 0}$ as $x_2 \to \infty$. If the solution is assumed to take the form
\be {\bm u}={\bm w}(k x_2) {\rm e}^{\ii k x_1},\la{normal} \en
where $k$ is the wave number and ${\bm w}$ is to be determined, then solving the above buckling problem yields the bifurcation condition in the implicit form
\be \omega(\lambda, kh)=0, \la{bifff} \en
where the expression for $\omega$ is obtained with the aid of {\it Mathematica}. Under the assumption that $r \ll 1$, we expect $kh$ and $1-\lambda$ to be both small. Writing $r= (kh)^3 r_0$ with $r_0$ an $O(1)$ constant, and looking for an asymptotic solution of the form
\be 1-\lambda=\xi_1 (kh)^2+ \xi_2 (kh)^3+\xi_3 (kh)^4+ \cdots, \la{feb100} \en
we may find the coefficients $\xi_1, \xi_2, ...$ by substituting \rr{feb100} into \rr{bifff} and equating the coefficients of $kh$ as successive orders. This yields the  asymptotic expansion \rr{0.3}.

\section{Derivation of the reduced model}
\setcounter{equation}{0}
We now proceed to derive a refined model that will yield the same asymptotic expansion \rr{0.3} without having to solve the full 3D problem. We have used both the procedure outlined in \cite{cai-fu2000} and the expansion method pioneered by \cite{ds2014} to derive the same equations. We shall present our derivations corresponding to the latter method since they are more compact than those obtained using the former method. We also wish to promote the use of the latter method since this method was often explained and used in more involved situations than the current one and as a result, its full potential does not seem to have been fully appreciated in the wider community.
\subsection{Refined model for the coating layer}
Following the strategy proposed by \cite{ds2014}, we first look for a series solution of the form
\be
u_1= \sum^{\infty}_{n=0} \frac{1}{n!}    U_n(x_1) x_2^n, \;\;\;\;\;
u_2= \sum^{\infty}_{n=0} \frac{1}{n!}   V_n(x_1) x_2^n, \la{ap1} \en
where the coefficient functions $ U_0(x_1), V_0(x_1), U_1(x_1), V_1(x_1), $ etc are to be determined. We note that the leading terms $U_0(x_1)$ and $V_0(x_1)$ are the displacement components at the traction-free surface $x_2=0$. Furthermore, the expansions in (\ref{ap1}) ensure that the traction and displacement conditions on the boundary $x_2=0$ are exactly satisfied.

On substituting \rr{ap1} into the traction-free boundary condition ${\bm t}(x_1, 0)={\bm 0}$ with the two components $T_1$ and $T_2$ given by \rr{tract}, we obtain
\be
U_1(x_1)=- V_0'(x_1), \;\;\;\;
V_1(x_1)=-({\cal A}_{1122}/{\cal A}_{2222}) U_0'(x_1). \la{ap2} \en
On the other hand, on substituting \rr{ap1} into \rr{equi1} and \rr{equi2} and equating the coefficients of like powers of $x_2^n$, we obtain the recurrence relations
\be
{\cal A}_{2121} U_{n}(x_1)=-({\cal A}_{1122}+{\cal A}_{1221})   V_{n-1}'(x_1) -{\cal A}_{1111}   U_{n-2}''(x_1)=0, \la{ap3} \en
\be
{\cal A}_{2222} V_{n}(x_1)=-({\cal A}_{1122}+{\cal A}_{1221})  U_{n-1}'(x_1) -{\cal A}_{1212}   V_{n-2}''(x_1)=0, \la{ap4} \en
which are valid for $n=2,3, ...$.
It can then be seen that these two recurrence relations together with \rr{ap2} can be used to express the right hand sides of \rr{ap1} entirely in terms of the leading terms
$U_0(x_1)$, $V_0(x_1)$, and their derivatives. This process may be carried out to any desired order. As a result, the traction vector ${\bm t}(x_1, h)$ at the interface  may also be expressed in terms of $U_0(x_1)$, $V_0(x_1)$ and their derivatives to any desired order in $h$. These expressions have the following special structure:
\be
t_1(x_1,h)= h c_1 U_0^{(2)}+ h^2 c_2 V_0^{(3)}+ h^3 c_3 U_0^{(4)}+ h^4 c_4 V_0^{(5)}+\cdots, \la{feb1} \en
\be
t_2(x_1,h)= h d_1 V_0^{(2)}+ h^2 d_2 U_0^{(3)}+ h^3 d_3 V_0^{(4)}+ h^4 d_4 U_0^{(5)}+\cdots, \la{feb2} \en
where the superscript \lq\lq $(n)$\rq\rq $(n=2, 3, ...)$ denotes the $n$th-order derivative with respect to $x_1$ and
the constant coefficients $c_1, d_1,  c_2, d_2$ etc only depend on the elastic moduli. For instance, the leading order coefficients are given by
$$ c_1={\cal A}^2_{1122}/{\cal A}_{2222} -{\cal A}_{1111}, \;\;\;\;
d_1={\cal A}_{1221}^2/{\cal A}_{2121}-{\cal A}_{1212}. $$
Observe the special feature that the $n$th-order derivatives $U_0^{(n)}$ and $V_n^{(n)}$ in \rr{feb1} and \rr{feb2} are multiplied by $h^{n-1}$ ($n=2, 3, ...$).

As the next step in deriving the reduced model, we need to express the traction vector ${\bm t}(x_1, h)$  in terms of the displacement components at $x_2=h$ which, with the use of \rr{ap1}, are given by
\be
U(x_1) \equiv u_1(x_1,h)= \sum^{\infty}_{n=0} \frac{1}{n!}  U_n(x_1) h^n , \la{ap55} \en
 \be V(x_1)\equiv u_2(x_1,h)=\sum^{\infty}_{n=0} \frac{1}{n!}  V_n(x_1) h^n, \la{ap5} \en
 where the sign \lq\lq $\equiv$" defines the short notations $U(x_1)$ and $V(x_1)$ that will be the only variables in our reduced model. Note that the sums in these two expressions only contain $U_0(x_1), V_0(x_1)$ and their derivatives. To express ${\bm t}(x_1, h)$  in terms of $U(x_1)$ and $V(x_1)$ (and their derivatives), we now invert \rr{ap55} and \rr{ap5} to express $U_0(x_1)$, $V_0(x_1)$ in terms of $U(x_1)$ and $V(x_1)$.

Viewing the two expressions \rr{ap55} and \rr{ap5} as asymptotic expansions for $U$ and $V$ in terms of the small parameter $h$, we may invert them by looking for an asymptotic solution of the form
\be
U_0(x_1)=U(x_1)+ h f_1(x_1)+h^2 f_2(x_1)+\cdots, \;\;\;\;V_0(x_1)=V(x_1)+ h g_1(x_1)+h^2 g_2(x_1)+\cdots, \la{ap6} \en
where the unknown functions $f_1(x_1), g_1(x_1)$ etc can be obtained by substituting \rr{ap6} into \rr{ap55} and \rr{ap5} and equating the coefficients of like powers of $h$. It turns out that although the right hand sides of \rr{ap55} and \rr{ap5} contain derivatives of $f_1(x_1), g_1(x_1), ...$ and their derivatives, the determination of these unknown functions at successive orders only involves the solution of linear algebraic equations, and these unknown functions can all be expressed in terms of $U, V$ and their derivatives. For instance, equating the coefficients of $h$ and $h^2$, we obtain
\be
f_1(x_1)=V^{(1)}(x_1), \;\;\;\; g_1(x_1)=\frac{{\cal A}_{1122}}{{\cal A}_{2222}} U^{(1)}(x_1), \la{mar1} \en
\be
f_2(x_1)=\left(\frac{{\cal A}_{1111} {\cal A}_{2222}+{\cal A}_{1122} {\cal A}_{1221}-{\cal A}^2_{1122}}{2{\cal A}_{1221} {\cal A}_{2222}} \right)U^{(2)}(x_1), \la{mar2} \en
\be
g_2(x_1)=\left( \frac{{\cal A}_{1212} -{\cal A}_{1221}  +{\cal A}_{1122}}{ 2 {\cal A}_{2222}}\right) V^{(2)}(x_1). \la{mar3} \en
As a result, on substituting \rr{ap6} back into \rr{feb1} and \rr{feb2}, the traction vector ${\bm t}(x_1, h)$ can be expressed in terms of $U(x_1)$, $V(x_1)$ and their derivatives to any desired order in $h$. These expressions take the form
\be
t_1(x_1,h)= h  {c}_1 U^{(2)}+ h^2 \hat{c}_2 V^{(3)}+ h^3 \hat{c}_3 U^{(4)}+ h^4 \hat{c}_4 V^{(5)}+\cdots, \la{febb1a} \en
\be
t_2(x_1,h)= h  {d}_1 V^{(2)}+ h^2 \hat{d}_2 U^{(3)}+ h^3 \hat{d}_3 V^{(4)}+ h^4 \hat{d}_4 U^{(5)}+\cdots, \la{febb2a} \en
where the new constant coefficients $\hat{c}_2, \hat{d}_2,  \hat{c}_3, \hat{d}_3, ...$ are also only dependent on the elastic moduli. Again observe the special feature that the $n$th-order derivatives $U^{(n)}$ and $V^{(n)}$ in these two expressions are multiplied by $h^{n-1}$ ($n=2, 3, ...$). The above derivation is straightforward. Although the expressions for the higher order coefficients $\hat{c}_2, \hat{d}_2,  \hat{c}_3, \hat{d}_3, ...$ are quite involved, the derivation can easily be implemented on a symbolic manipulation platform such as {\it Mathematica}.

As pointed out by \cite{ds2014}, see also \cite{wsd2019}, the expansions such as \rr{febb1a} and \rr{febb2a} contain all the necessary terms that can be used to recover any existing plate theories although this has not been carried out for higher order plate theories such as the Timoshenko theory. We now use these expansions to derive our refined model.

We use $\hat{\cal A}_{jilk}$ to denote the moduli for the half-space. We assume that the thickness $h$, displacement $u_i$ and coordinates $x_i$  have all been scaled by a typical lengthscale $L$. For instance, for analysis of buckling, we shall take $L=1/k$ with $k$ denoting the wavenumber of the buckling mode. We assume the scaled $h$ to be small and, as indicated in the Introduction, consider the case when $\hat{\cal A}_{jilk}= O(h^3 {\cal A}_{jilk})$, that is, the layer is much stiffer than the half-space. It is this parameter regime in which the layer behaviour can be modeled by the Euler-Bernoulli beam theory to leading order.
In this case, the layer will deform like a beam in the sense that its vertical displacement will be much larger than its axial displacement; more precisely, $U=O(h V)$.
We expect that such a layer will buckle at small compressive strains, that is with $\lambda \approx 1$. Thus, we write
\be
\lambda=1+h^2 \psi, \la{psi} \en
where $\psi$ is a constant of $O(1)$. As a result, the moduli can be expanded as
\be {\cal A}_{jilk}=\left.{\cal A}_{jilk}\right|_{\lambda=1}+h^2 \psi {\cal A}'_{jilk}+\frac{1}{2} h^4 \psi^2 {\cal A}''_{jilk}+\cdots, \la{dmoduli} \en
where e.g. $ {\cal A}'_{jilk}$ denotes the derivative of $ {\cal A}_{jilk}$ with respect to $\lambda$ evaluated at $\lambda=1$. To simplify notation, we shall write
$\left.{\cal A}_{jilk}\right|_{\lambda=1}$ simply as $ {\cal A}_{jilk}$. Since $\lambda=1$ corresponds to the undeformed state which is isotropic, if we denote the associated Lame constants by $\lambda^*$ and $\mu$, we have
\be {\cal A}_{jilk}= \lambda^* \delta_{ji}\delta_{lk}+ \mu (\delta_{jl}\delta_{ik}+\delta_{jk}\delta_{il}). \la{lame} \en
The derivatives ${\cal A}'_{jilk}$ are dependent on the strain-energy function used and are not written out here for the sake of brevity.

We look for an asymptotic solution of the form
\be
U(x_1)=h w_{11}(x_1)+h^2 w_{12}(x_1)+h^3 w_{13}(x_1)+h^4 w_{14}(x_1)+\cdots, \la{feb222} \en
\be
V(x_1)=w_{20}(x_1)+h w_{21}(x_1)+h^2 w_{22}(x_1)+h^3 w_{23}(x_1)+\cdots, \la{ffeb2} \en
where the functions $w_{20}, w_{11}, w_{21}$ etc are to be determined. Since the elastic moduli of the half-space is of order $h^3$, the tractions exerted by the half-space can at most be order $h^3$. It turns out that the vertical traction is indeed of order $h^3$, but the horizontal traction can only be of order $h^4$ for asymptotic consistency \citep{cai-fu2000}. This of course is why the Winkler assumption is valid to leading order.
Thus, we write traction continuity in the form $$t_1(x_1, h)=h^4 \hat{t}_1(x_1),\;\;\;\;  t_2(x_1, h)=h^3 \hat{t}_2(x_1),$$ with $h^4 \hat{t}_1(x_1)$ and $h^3 \hat{t}_2(x_1)$ denoting the tractions exerted by the half-space. As a result, \rr{febb1a} and \rr{febb2a} may be replaced by
\be
h^4 \tilde{t}_1(x_1) = h  {c}_1 U^{(2)}+ h^2 \hat{c}_2 V^{(3)}+ h^3 \hat{c}_3 U^{(4)}+ h^4 \hat{c}_4 V^{(5)}+\cdots, \la{febb1} \en
\be
h^2 \tilde{t}_2(x_1) =   {d}_1 V^{(2)}+ h \hat{d}_2 U^{(3)}+ h^2 \hat{d}_3 V^{(4)}+ h^3 \hat{d}_4 U^{(5)}+\cdots, \la{febb2} \en
where we have divided \rr{febb2a} by $h$ to ease our presentation.

On substituting \rr{feb222} and \rr{ffeb2} together with \rr{psi} and \rr{dmoduli}  into \rr{febb1} and \rr{febb2} and equating the coefficients of like powers of $h$, we can find $w_{11}, w_{21}$ etc at successive orders in terms of the traction functions $\tilde{t}_1(x_1)$ and $\tilde{t}_2(x_1)$. The two equations at order $h$ are automatically satisfied, whereas the two equations at order $h^2$ are linear algebraic equations and can be solved to give
\be
w_{11}^{(2)}=-\frac{1}{2} w_{20}^{(3)}, \la{feb4} \en
 \be
   \frac{\mu (\lambda^*+\mu)}{3 (\lambda^*+2 \mu)}  w_{20}^{(4)}-
\psi \bar{\sigma}_1' w_{20}^{(2)} =\tilde{t}_2(x_1), \la{feb6} \en
where $\bar{\sigma}_1$ has been defined in (\ref{connect}) and  $\bar{\sigma}_1'$ stands for the derivative of $\bar{\sigma}_1$ with respect to $\lambda$ evaluated at $\lambda=1$. The leading order result \rr{feb6} corresponds to an Euler-Bernoulli beam that is only supported by normal forces.

At order $h^3$, the two linear algebraic equations can be solved to give
\be w_{12}^{(2)}=-\frac{1}{2} w_{21}^{(3)}, \la{feb5} \en
\be
 \frac{\mu (\lambda^*+\mu)}{3 (\lambda^*+2 \mu)}  w_{21}^{(4)}-
\psi \bar{\sigma}_1' w_{21}^{(2)} = 0. \la{feb8} \en
At order $h^4$, we solve the two equations to obtain
\be w_{13}^{(2)}=-\frac{1}{2} w_{22}^{(3)}-\frac{(\lambda^* +\mu )}{6
   (\lambda^* +2 \mu )}w_{20}^{(5)}+\frac{\lambda^*\bar{\sigma}_1'  \psi}{8\mu(\lambda^* + \mu )}w_{20}^{(3)}-
\frac{(\lambda^* +2 \mu )}{4 \mu  (\lambda^* +\mu
   )} \tilde{t}_1(x_1), \la{feb7} \en
\be
\frac{\mu (\lambda^*+\mu)}{3 (\lambda^*+2 \mu)}  w_{22}^{(4)}-
\psi \bar{\sigma}_1' w_{22}^{(2)} = \frac{1}{2} \tilde{t}_{1}'(x_1) -\frac{\mu  (\lambda^* +\mu )}{15 (\lambda^* +2 \mu )} w_{20}^{(6)}+ k_4 \psi w_{20}^{(4)}+\frac{1}{2} \bar{\sigma}_{1}'' \psi^2 w_{20}^{(2)}, \la{feb9} \en
where
$$ k_4=-\frac{1}{12 (\lambda^* +2 \mu )^2} \left\{ 2 \lambda^* (\lambda^*+2\mu)\bar{\sigma}_{1}'
     +  (\lambda^{*} +2 \mu)^2   {\cal A}'_{1111}   -2 \lambda^{*}(\lambda^{*}+2 \mu)
    {\cal A}'_{1122}+\lambda^{*2} {\cal A}'_{2222} \right\}. $$


Multiplying \rr{feb8} and \rr{feb9} by $h$ and $h^2$ , respectively, and adding the two resulting equations to \rr{feb6}, we obtain, after making use of \rr{ffeb2},
$$
\frac{\mu (\lambda^*+\mu)}{3 (\lambda^*+2 \mu)}  V^{(4)}-
\psi \bar{\sigma}_1' V^{(2)} =  \tilde{t}_{2}(x_1)+\frac{h^2}{2 } \tilde{t}_{1}'(x_1)+\frac{1}{2} \bar{\sigma}_{1}'' \psi^2 h^2 V^{(2)} $$
\be
+ k_4 h^2 \psi V^{(4)}-\frac{\mu  (\lambda^* +\mu )}{15 (\lambda^* +2 \mu )} h^2 V^{(6)}+O(h^3). \la{feb10} \en
 On the other hand, multiplying \rr{feb4} and \rr{feb5} by $h^2$ and $h^3$, respectively, and adding the two resulting equations together, we obtain, after making use of \rr{feb222} and rearranging,
$$
h^4 \tilde{t}_1(x_1)=-\frac{4 \mu  (\lambda^* +\mu )}{\lambda^* +2 \mu } \left(h U^{(2)}(x_1)+\frac{1}{2} h^2
   V^{(3)}(x_1)\right) -\frac{2   \mu  (\lambda^* +\mu )^2 }{3 (\lambda^* +2\mu )^2} h^4 V^{(5)}(x_1) $$
   \be
   + \frac{\lambda^*\bar{\sigma}_1'}{2(\lambda^*+2\mu)}h^2 (\lambda-1) V^{(3)}(x_1)+O(h^5),
   \la{feb11} \en
where we have used \rr{psi} to eliminate $\psi$.
It can be shown that the same equations for $U$ and $V$ are obtained even after solutions at the next order are incorporated. Thus, the error terms in \rr{feb10} and \rr{feb11} may be replaced by $O(h^4)$ and $O(h^6)$, respectively.

Finally, on eliminating $\tilde{t}_1(x_1)$ from \rr{feb10} with the use of \rr{feb11} and solving the resulting equation for $t_2(x_1)$, we obtain
$$
  h^3 \tilde{t}_2(x_1)=\frac{\mu (\lambda^*+\mu)}{3 (\lambda^*+2 \mu)}  h^3 V^{(4)}-
 (\lambda-1)  h \bar{\sigma}_1' V^{(2)}+\frac{2 \mu (\lambda^*+ \mu)}{\lambda^*+2 \mu} \left( h^2 U^{(3)}  + \frac{1}{2} h^3 V^{(4)} \right)  $$
 \be +\frac{h^5 \mu  \left(6 \lambda^{*2}+13 \lambda^*  \mu +7 \mu
   ^2\right)}{15 (\lambda^* +2 \mu )^2} V^{(6)}  + \hat{k}_4 h^3 (\lambda-1) V^{(4)}-\frac{1}{2} h (\lambda-1)^2 \bar{\sigma}_1'' V^{(2)}+O(h^7), \la{feb12} \en
 where
 $$ \hat{k}_4= \frac{1}{12 (\lambda^* +2 \mu )^2} \left\{-\lambda^* (\lambda^*+2\mu)\bar{\sigma}_{1}'
     +  (\lambda^{*} +2 \mu)^2   {\cal A}'_{1111}   -2 \lambda^{*}(\lambda^{*}+2 \mu)
    {\cal A}'_{1122}+\lambda^{*2} {\cal A}'_{2222} \right\}. $$
Note that the first two terms on the right hand side of \rr{feb12} corresponds to the leading order theory, and we have not combined them with the other higher order terms. In particular, although the second term involving $h^3 V^{(4)}$ is of the same order as the leading terms, the sum $h^2 U^{(3)}  + \frac{1}{2} h^3 V^{(4)}$ together is of $O(h^5)$, as can be seen from \rr{feb11} where all terms must necessarily be of the same order.

\subsection{Response of the half-space}
Having derived the relation between traction and displacement at the interface from the side of the coating layer, we now employ the Stroh formulation \citep{stroh1958} to derive its counterpart from the side of half-space. The following derivations are adapted from \cite{fu2007}.

The equilibrium equations for the half-space are
$$
\hat{\cal A}_{1i1 k} \hat{u}_{k,11}+(\hat{\cal A}_{1i2 k}+\hat{\cal A}_{2i1 k}) \hat{u}_{k,12}+\hat{\cal A}_{2i1 2} \hat{u}_{k,22}=0, $$
or equivalently
\be Q \hat{\bm u}_{,11}+(R+R^T)  \hat{\bm u}_{,12}+T \hat{\bm u}_{,22}={\bm 0}, \la{equi19} \en
where the three matrices $T, R$ and $Q$ are defined by
\be T_{ik}=\hat{\cal A}_{2i2k}, \;\;\;\; R_{ik}=\hat{\cal A}_{1i2k}, \;\;\;\;
Q_{ik}=\hat{\cal A}_{1i1k}. \la{1.7} \en
The traction vector $\hat{\bm t}$ on any surface with unit normal ${\bm n}=(\delta_{i2})$ is given by
\be
\hat{t}_i =\hat{\chi}_{i2}=\hat{\cal A}_{2il k} \hat{u}_{k,l}=\hat{\cal A}_{2i1 k} \hat{u}_{k,1}+\hat{\cal A}_{2i2 k} \hat{u}_{k,2},
\la{2.166}
\en
or equivalently,
\be \hat{\bm t}=R^T \hat{\bm u}_{, 1}+T \hat{\bm u}_{,2}. \la{tract10} \en

We define the Fourier transform (FT) of $\hat{\bm u}$ and its inverse by
\be
{\bm z}(k, x_2)\equiv {\cal F}[\hat{\bm u}] =\int_{-\infty}^{\infty} \hat{\bm u}(x_1, x_2) {\rm e}^{-\ii k x_1} dx_1, \la{FT0} \en
\be
 \hat{\bm u}(x_1, x_2)=\frac{1}{2 \pi} \int_{-\infty}^{\infty} {\bm z}(k, x_2) {\rm e}^{\ii k x_1} dk, \la{FT} \en
where the first expression defines the vector function ${\bm z}(k, x_2)$.
Applying the FT to \rr{equi19}, we obtain
\be T {\bm z}''+\ii k (R+R^T)  {\bm z}'-k^2 Q {\bm z}={\bm 0}, \la{equi1a} \en
where a prime denotes differentiation with respect to $x_2$. On substituting a trial solution of the form
\be
{\bm z}(k, x_2) ={\bm a}\,\ee^{ \ii k p x_2}, \la{equilb} \en
into \rr{equi1a},  we find that the constant scalar $p$ and vector ${\bm a}$
satisfies the eigenvalue problem \be \left(p^2 \t+p
(\r+\r^T)+\q\right) {\bm a}=\o. \la{1.6} \en
Under the assumption that  $\hat{\cal A}_{jilk}$ satisfies the strong ellipticity condition, the
eigenvalues of $p$ in \rr{1.6} cannot be pure real. For the current plane-strain problem, we denote by $p^{(1)}, p^{(2)}$ the two eigenvalues of $p$ with positive imaginary
parts and ${\bm  a}^{(1)}, {\bm a}^{(2)}$ the
associated eigenvectors. Then for $k>0$ a general solution that satisfies
the decaying condition is \be{\bm z}(k, x_2)=
\sum_{j=1}^{2} c_j {\bm a}^{(j)} \ee^{ \ii k p^{(j)}
x_2}  =A \langle \ee^{ \ii k
p x_2} \rangle \,{\bm c}=A \langle \ee^{ \ii k
p x_2} \rangle A^{-1} \,{\bm d}, \la{1.8} \en
where $c_1, c_2$ are constants, $$ A=[{\bm a}^{(1)}, {\bm
a}^{(2)}], \;\; {\bm c}=[c_1, c_2]^T, $$ ${\bm d}=A {\bm c}$, and
$\langle \ee^{ \ii k p x_2} \rangle$ denotes the diagonal
matrix $$ {\rm diag}\,\{\ee^{ \ii k p^{(1)} x_2}, \ee^{ \ii
k p^{(2)} x_2}\}.$$
For $k<0$, we must choose the conjugates of $p^{(1)}, p^{(2)}$ in constructing the general solution in order to satisfy the decaying condition. Thus, we have
 \be{\bm z}(k, x_2)=\bar{A} \langle \ee^{ \ii k
\bar{p} x_2} \rangle \bar{A}^{-1} \,\bar{\bm d},\;\;\;\; k<0, \la{1.888} \en
where an overbar signifies complex conjugation.
It then follows that
 \be {\bm z}(k, x_2)=\overline{{\bm z}(-k, x_2)},\;\;\;\; k<0. \la{negk} \en
With the aid of \rr{1.8} and \rr{1.888}, we deduce that \be
 {\bm z}'=\ii k \cdot
\left\{
\begin{array}{ll}  A \langle
p   \rangle \,A^{-1} {\bm z}(k, x_2), & {\rm when}\, k>0,
\\  \bar{A} \langle   \bar{p}   \rangle
\,\bar{A}^{-1} {\bm z}(k, x_2), & {\rm when}\, k<0.
\end{array} \right. \la{dzdx2}\en
Thus,  we have \be  {\bm z}' = \left({\rm Re}\,G+ \ii
\,{\rm sgn}(k)\,{\rm Im}\,G \right)\, {\cal F}\left[
\frac{\paa \hat{\bm u}}{\paa x_1}\right], \la{1.19} \en
where  \be G=A\langle p \rangle A^{-1}, \la{1.18} \en  Re and Im
denote the real and imaginary parts, respectively, and sgn is the
sign function.

To invert \rr{1.19}, we make use of the result that
\be
{\rm sgn}
(k)=-\frac{1}{\ii \pi} {\cal F}[\frac{1}{x_1}], \la{1.22} \en
where the Fourier transform is understood to take its principal value.
It then follows by inverting \rr{1.19}, followed by the use of the convolution theorem, that
\be
 \hat{\bm u}_{,2}=  ({\rm
Re}\,G)\hat{\bm u}_{,1}-({\rm Im}\,G) \,
\frac{1}{\pi x_1}*\hat{\bm u}_{,1}, \la{1.20a} \en
where the star  signifies integral convolution. Alternatively, we may write this result as
 \be
  \hat{\bm u}_{,2}=  ({\rm
Re}\,G)\hat{\bm u}_{,1}+ ({\rm Im}\,G) \,
{\cal H}[\hat{\bm u}_{,1}], \la{1.20} \en
where ${\cal H}$ denotes the Hilbert transform defined by \be
{\cal H}[g(x_1)]=\frac{1}{\pi}\,{\rm p.v.} \int^{\infty}_{-\infty}
\frac{g(y)}{y-x_1} dy=-\frac{1}{\pi x_1} \star g(x_1).
\la{1.21} \en
On eliminating $\hat{\bm u}_{,2}$ from \rr{tract10} with the use of \rr{1.20}, we obtain
 \be \hat{\bm t}=\left\{R^T+ T \,({\rm
Re}\,G)\right\}   \hat{\bm u}_{, 1} + T({\rm Im}\,G) \,
{\cal H}[ \hat{\bm u}_{,1}]. \la{1.9} \en
Note that the continuity condition $\hat{\bm u}(x_1, h)= {\bm u}(x_1, h)$ at the interface may be differentiated to yield
$\hat{\bm u}_{,1}(x_1, h)= {\bm u}_{,1}(x_1, h)$. As a result, when evaluated at the interface $x_2=h$, the $\hat{\bm u}$ in \rr{1.9} may be replaced by
${\bm u}(x_1, h)=\{ U(x_1), V(x_1)\}^T$.

It was shown in \cite{fu2007} that the matrix $G$ is related to the surface impedance matrix $M$ \citep{it1969} by
\be M=-\ii  (R^T+T G), \la{impp} \en
so that equation \rr{1.9} may also be rewritten as
\be \hat{\bm t}= - ({\rm
Im}\,M)   \hat{\bm u}_{, 1}+ ({\rm Re}\,M) \,
{\cal H}[ \hat{\bm u}_{,1}]. \la{1.91} \en
The surface impedance matrix $M$ plays an important role in the surface wave theory \citep{bl1985} and many useful results about $M$ are known. In particular, it is positive definite under the convexity assumption for the elastic moduli, and satisfies the matrix Riccati equation
\be
(M-\ii R) T^{-1} (M+\ii R^{\rm T})-Q=0. \la{add7} \en
See \citet{bir1985}, \citet{fu-mielke2002}. This matrix equation can be solved analytically when there are enough symmetries in the problem.   For the current plane-strain problem, the three matrices $Q, R, T$ have the simple forms
\be \t=\left(\begin{array}{cc} T_1 & 0 \\ 0 & T_2 \end{array} \right),
\;\;\;\;
\r=\left(\begin{array}{cc} 0 & R_1 \\ R_2 & 0 \end{array} \right),
\;\;\;\;
\q=\left(\begin{array}{cc} Q_1 & 0 \\ 0 & Q_2 \end{array} \right), \la{1.16}
\en
where
$$ T_1=\hat{\cal A}_{2121}, \;\; T_2=\hat{\cal A}_{2222},\;\;
R_1=\hat{\cal A}_{1122}, $$ $$  R_2=\hat{\cal A}_{2112}, \;\;
Q_1=\hat{\cal A}_{1111}, \;\; Q_2=\hat{\cal A}_{1212}.  $$
As a result, equation \rr{add7} can be solved explicitly to yield
\be
M=\left(\begin{array}{cc} M_1 & \ii M_4 \\
-\ii M_4 & M_2
\end{array} \right), \;\;\;\;
M_i \; {\rm real}, \la{may2} \en
with
$$
M_1=\sqrt{T_1 Q_1-\frac{T_1}{T_2} \left(
\frac{R_1+R_2}{1+\gamma} \right)^2}, \;\;\;\; \gamma=\sqrt{\frac{T_1 Q_2}{T_2 Q_1}},$$
\be
M_2=\gamma \frac{T_2}{T_1} M_1, \;\;\;\;
 M_4=\frac{\gamma R_1-R_2}{1+\gamma}, \la{may3} \en \medskip
Returning to \rr{1.91} and evaluating it at the interface $x_2=h$, we may replace $\hat{\bm t}$ by $\{ h^4 \tilde{t}_1, h^3 \tilde{t}_2 \}^T$ and $\hat{\bm u}$ by $\{ U, V \}^T$  to obtain
\be
-M_4 V^{(1)}(x_1)+M_1 {\cal H}[ U^{(1)}(x_1)]= h^4 \tilde{t}_1,
\la{u2dd1b} \en
\be
M_4 U^{(1)}(x_1)+M_2 {\cal H}[ V^{(1)}(x_1)]=h^3 \tilde{t}_2. \la{amp1b} \en
With $h^4 \tilde{t}_1$ and $h^3 \tilde{t}_2$ given by \rr{feb11} and \rr{feb12}, equations \rr{u2dd1b} and \rr{amp1b} are two ordinary differential equations for $U(x_1)$ and $V(x_1)$, and are the refined film/substrate model that we set out to derive.

The classical model corresponds to balancing the term $M_2 {\cal H}[ V^{(1)}(x_1)]$ with the first two terms in $h^3 \hat{t}_2$ and is given by
\be M_2 {\cal H}[ V^{(1)}(x_1)]=\frac{\mu (\lambda^*+\mu)}{3 (\lambda^*+2 \mu)}  h^3 V^{(4)}-
 (\lambda-1)  h \bar{\sigma}_1' V^{(2)}. \la{fff1} \en
For the classical model, $M_2$ may be evaluated at $\lambda=1$ and we have
 \be M_2=\frac{2 \mu_{\rm s}(\lambda_s+2 \mu_{\rm s})}{\lambda_s+3 \mu_{\rm s}}, \;\;\;\; \bar{\sigma}_1'=\frac{4 \mu (\lambda^*+\mu)}{\lambda^*+2 \mu}, \la{fff2} \en
 where $\lambda_s$ and $\mu_{\rm s}$ represent the Lame constants for the substrate. For sinusoidal solutions of the form $V(x_1)=A \sin (kx_1)$ with $A$ a constant, we have
$$ {\cal H}[ V^{(1)}(x_1)]=A k {\cal H}[\cos (kx_1)]=-A k \sin (kx_1)=-k V(x_1). $$
Equation \rr{fff1} then recovers \rr{classical}.

\section{Self-consistency of the refined model}
\setcounter{equation}{0}
On substituting a periodic buckling solution of the form
\be U(x_1)=A \sin (kx_1), \;\;\;\; V(x_1)=B \cos (kx_1) \la{4.1} \en
into the reduced model \rr{u2dd1b} and \rr{amp1b} and canceling $\sin (kx_1)$ and $\cos (kx_1)$, we obtain two homogeneous linear equations for $A$ and $B$. For a non-trivial solution, we must set the determinant of the coefficient matrix to zero, which then yields the bifurcation condition. By assuming that $kh$ is small, and both $ {\lambda}_s/\lambda^*$ and ${\mu}_s/\mu$ are of order $(kh)^3$, we find from the bifurcation condition that $1-\lambda$ has an asymptotic expansion given by
$$
1-\lambda=\frac{2 (\nu-1)^2}{3-4 \nu} \frac{r}{k h} +\frac{1}{12} (kh)^2+d_0 r+d_1 (k h)^4+d_2 r (k h) $$
\be
+d_3 \frac{ r^2}{ (kh)^2}+d_4 (kh)^2 r+d_5 \frac{r^2}{kh}+O((kh)^6), \la{app1} \en
where the constants $d_0, d_1, d_2$ and $d_3$ are the same as in \rr{0.3}, and the new constants $d_4$ and $d_5$ are given by
\be
d_4=\frac{192 \nu ^4-512 \nu ^3+508 \nu
   ^2-218 \nu +33}{24 (4 \nu -3)^3}, \la{app2a} \en \be d_5=\frac{256 \nu ^6-992 \nu ^5+1608 \nu
   ^4-1380 \nu ^3+644 \nu ^2-147 \nu
   +11}{2 (4 \nu -3)^3}. \la{app2} \en
Compared with \rr{0.3}, the expansion \rr{app1} contains two extra terms of order $(kh)^5$. We have checked to verify that the expansion \rr{app1} is the same as that given by the exact bifurcation condition, thus verifying the self-consistency of the refined model. We now show that the two extra  order $(kh)^5$ terms in \rr{app1} can significantly improve its accuracy.

The maximum stretch is attained when $d \lambda/d (kh)=0$. On substituting \rr{app1} and
\be
kh=g_1 r^{1/3}+ g_2 r^{3/3}+ g_3 r^{4/3}+ O(r^{5/3}) \la{kh} \en
into this equation and then equating the coefficients of like powers of $r$ to zero, we obtain
\be
g_1=2^{2/3} \sqrt[3]{3}
   \sqrt[3]{\frac{(\nu
   -1)^2}{3-4 \nu }}, \;\;\;\; g_2=\frac{-1920 \nu ^4+3776 \nu
   ^3-2504 \nu ^2+500 \nu
   +33}{60 (3-4 \nu )^2}, \la{app3} \en \be
    g_3=\frac{256 \nu ^5-640 \nu
   ^4+592 \nu ^3-208 \nu ^2-2
   \nu +11}{2\ 3^{2/3} (3-4
   \nu )^2 \sqrt[3]{8 \nu
   ^2-14 \nu +6}}. \la{app3a} \en
On substituting the expansion \rr{kh}  back into \rr{app1}, we obtain the corresponding expression for the critical stretch
\be
\lambda_{\rm cr}=1-\frac{3^{2/3} (1-\nu
   )^{4/3}}{(6-8 \nu )^{2/3}} r^{2/3}+\frac{(\nu -1) (2 \nu -1)}{4
   \nu -3} r + g_4 r^{4/3} +g_5 r^{5/3}+O(r^{6/3}), \la{app4} \en
where
\be g_4= \frac{(1-\nu )^{2/3} \left(480
   \nu ^4-556 \nu ^3+199 \nu
   ^2-115 \nu +72\right)}{20\
   3^{2/3} \sqrt[3]{6-8 \nu }
   (3-4 \nu )^2}, \la{app5a} \en \be
    g_5=-\frac{\sqrt[3]{3-4 \nu }
   \left(256 \nu ^6-1040 \nu
   ^5+1796 \nu ^4-1670 \nu
   ^3+863 \nu ^2-227 \nu
   +22\right)}{2\ 2^{2/3}
   \sqrt[3]{3} (1-\nu )^{2/3}
   (4 \nu -3)^3}. \la{app5} \en
In the incompressibility limit $\nu \to 0.5$, the expressions for the critical stretch and wavenumber reduce to
\be
\lambda_{\rm cr}=1-\frac{1}{4} (3r)^{2/3}+\frac{11}{160} (3r)^{4/3}-\frac{1}{24} (3r)^{5/3} +O(r^2), \la{app6} \en
\be
(kh)_{\rm cr}= (3r)^{1/3}+\frac{3}{20} r+O(r^{5/3}). \la{app7} \en
Since care needs to be taken in using the truncated asymptotic expansion \rr{app1} to compute $d \lambda/d (kh)$,
we have checked to verify that all the terms displayed in \rr{app3}$-$\rr{app7} are the same as those given by the exact theory, and those remainder terms represented by the $O$ symbol cannot be derived using the current reduced model.

We observe that in the incompressible case, we have $g_3=0$ so that adding the $O((kh)^5)$ terms in \rr{app1} does not result in any change in $(kh)_{\rm cr}$ but does give rise to an extra $O(r^{5/3})$ term in \rr{app6}. In Fig.1(a,b), we have shown the effect of adding the $O(r^{5/3})$ term in \rr{app6} by comparing the asymptotic results with their exact counterparts. It is seen that adding the $O(r^{5/3})$  term in \rr{app6} significantly improves the accuracy of the resulting asymptotic expansion for $\lambda_{\rm cr}$.

\begin{figure}[ht]
\begin{center}
\begin{tabular}{ccc}
\includegraphics[scale=0.4]{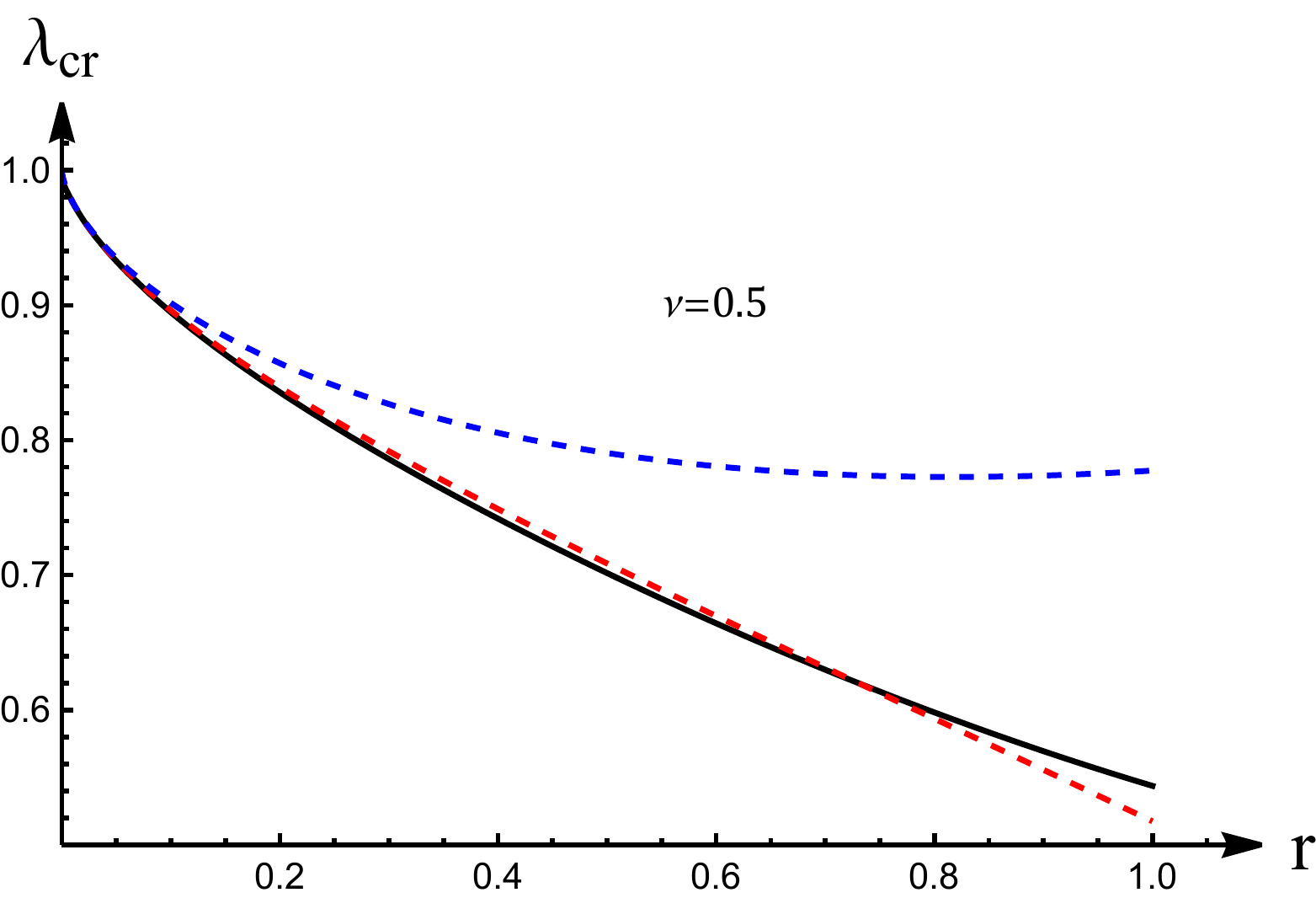}& & \includegraphics[scale=0.4]{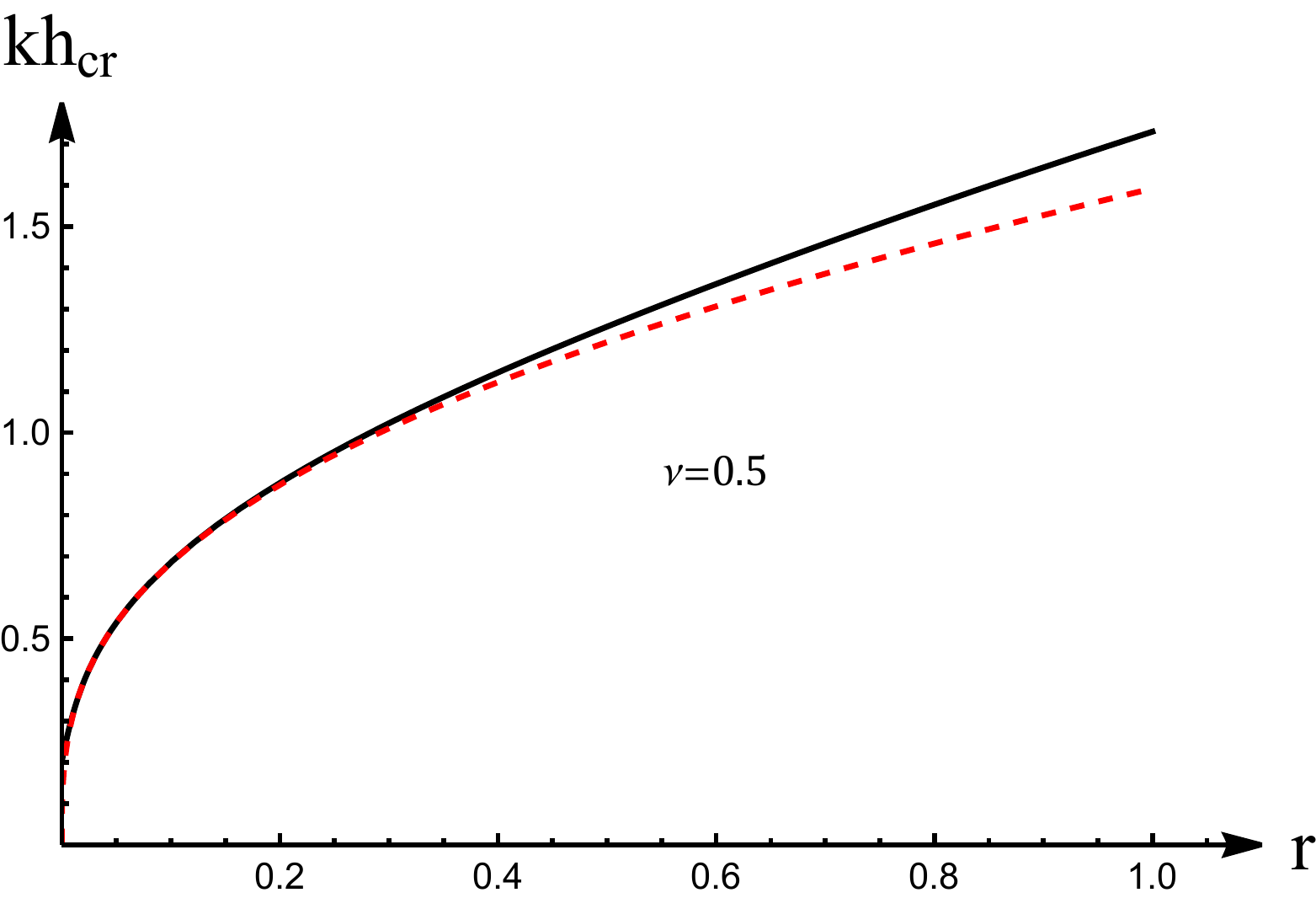} \\
(a) & & (b)
\end{tabular}
\caption{Comparison of asymptotic results (dashed lines) with exact results (black solid line) in the incompressible case. The red dashed lines in (a) and (b) correspond to \rr{app6} and \rr{app7}, respectively, and the blue line in (a) corresponds to \rr{app6} with the $O(r^{5/3})$ term neglected.}
\label{h1}
\end{center}
\end{figure}

Having demonstrated the importance of the extra $O((kh)^5)$ terms in \rr{app1}, we now investigate their effect for the compressible case. We note that the extra $O((kh)^5)$ terms in \rr{app1} give rise to the $g_3 r^{4/3}$ term in \rr{kh} and the $g_5 r^{5/3}$ term in \rr{app4}. In Fig.~2 to Fig.~4, we have shown the counterparts of Fig.~1 for $\nu=0.45, 0.3$ and $0.1$, respectively, where the red and blue dashed lines are asymptotic results with the these extra terms added or neglected, respectively. It is seen that adding the extra terms continues to improve the accuracy for $\nu$ close to $0.5$ (e.g. $\nu=0.45$), but it has the opposite effect when $\nu=0.3$ and a mixed effect when $\nu=0.1$ (in the sense that it improves the accuracy in $\lambda_{\rm cr}$ but worsens the accuracy in $(kh)_{\rm cr}$). Thus, we conclude that including the $O((kh)^5)$ terms in \rr{app1} improves the accuracy of the asymptotic results significantly only when the materials are incompressible or nearly incompressible. For other values of $\nu$, the optimal truncation of the asymptotic expansion \rr{app1} will be dependent on $\nu$.

  \begin{figure}[ht]
\begin{center}
\begin{tabular}{ccc}
\includegraphics[scale=0.35]{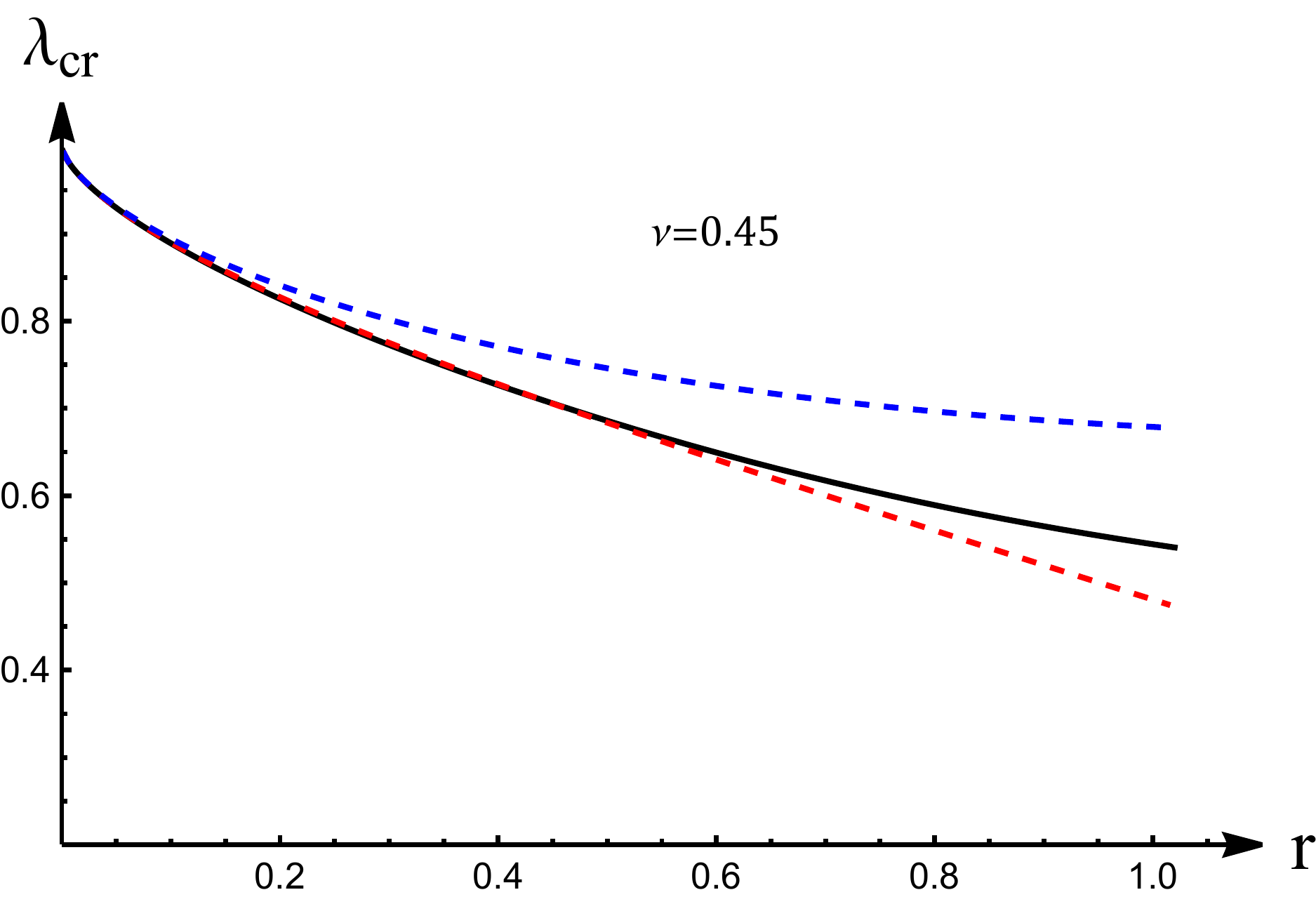}& & \includegraphics[scale=0.35]{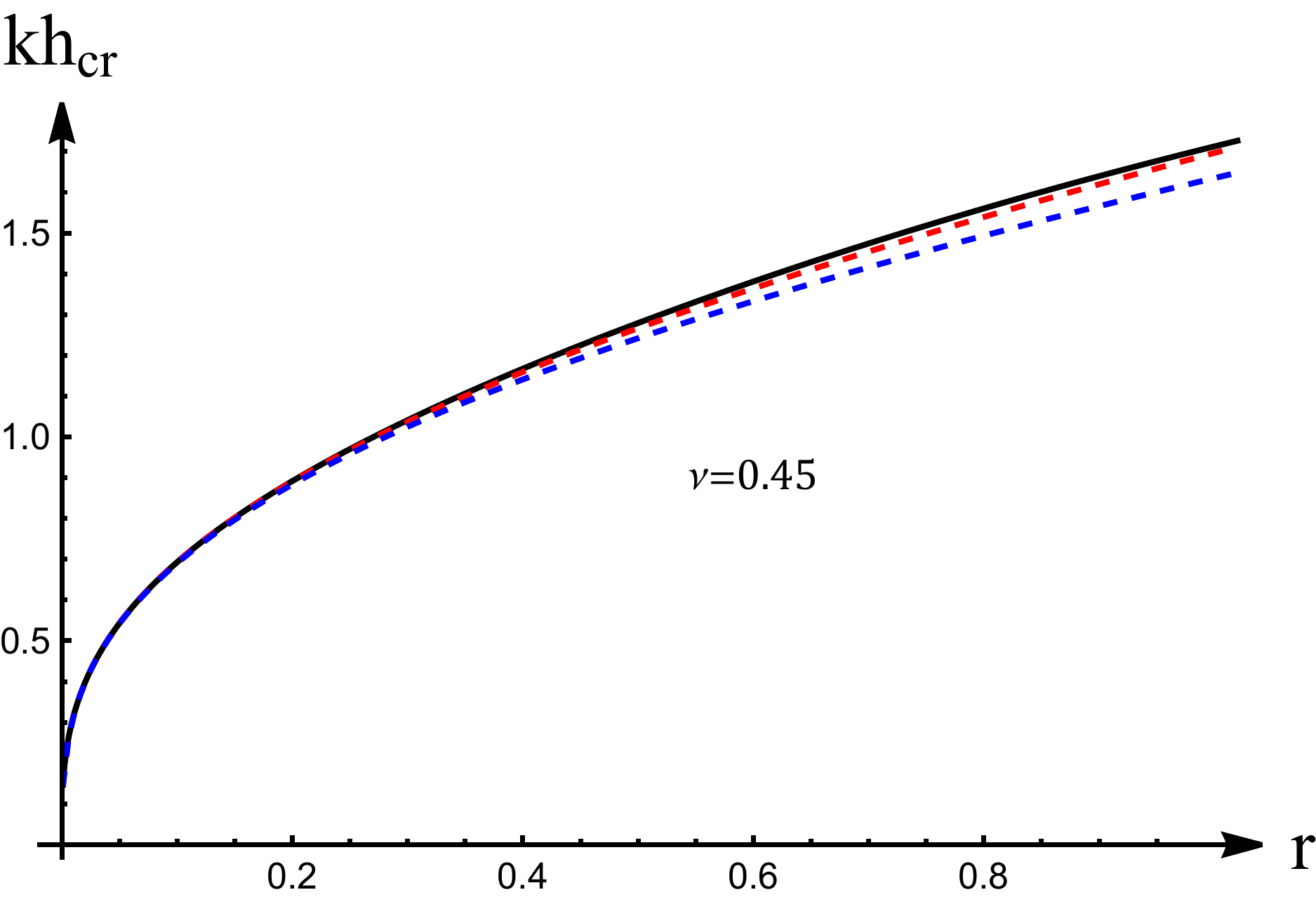} \\
(a) & & (b)
\end{tabular}
\caption{Comparison of asymptotic results (dashed lines) with exact results (black solid line) when $\nu=0.45$. The red dashed lines in (a) and (b) correspond to \rr{app4} and \rr{kh}, respectively, and whereas the blue lines represent lower order approximations with $g_3$ and $g_5$ set to zero.}
\label{h1}
\end{center}
\end{figure}

 \begin{figure}[ht]
\begin{center}
\begin{tabular}{ccc}
\includegraphics[scale=0.35]{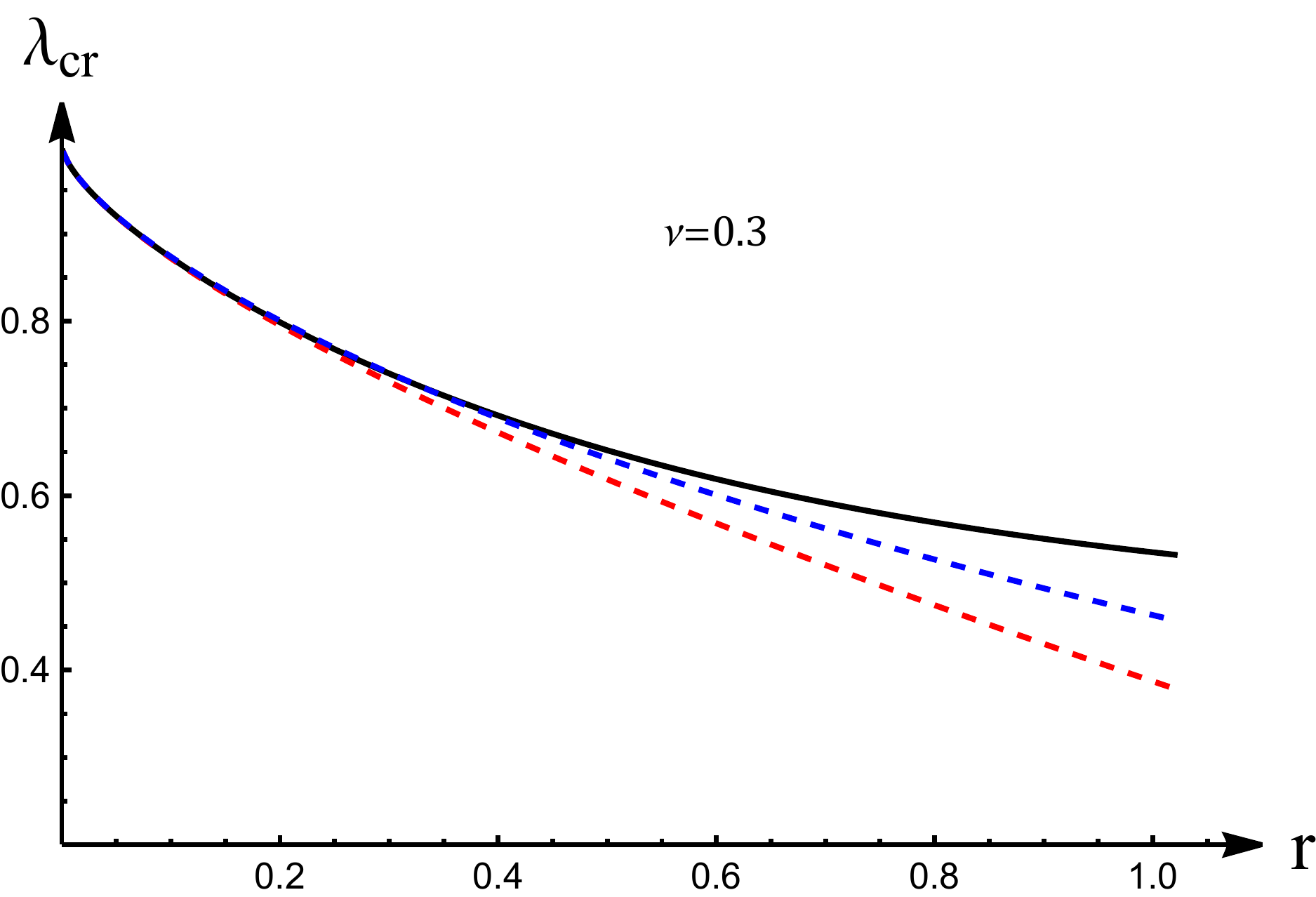}& & \includegraphics[scale=0.35]{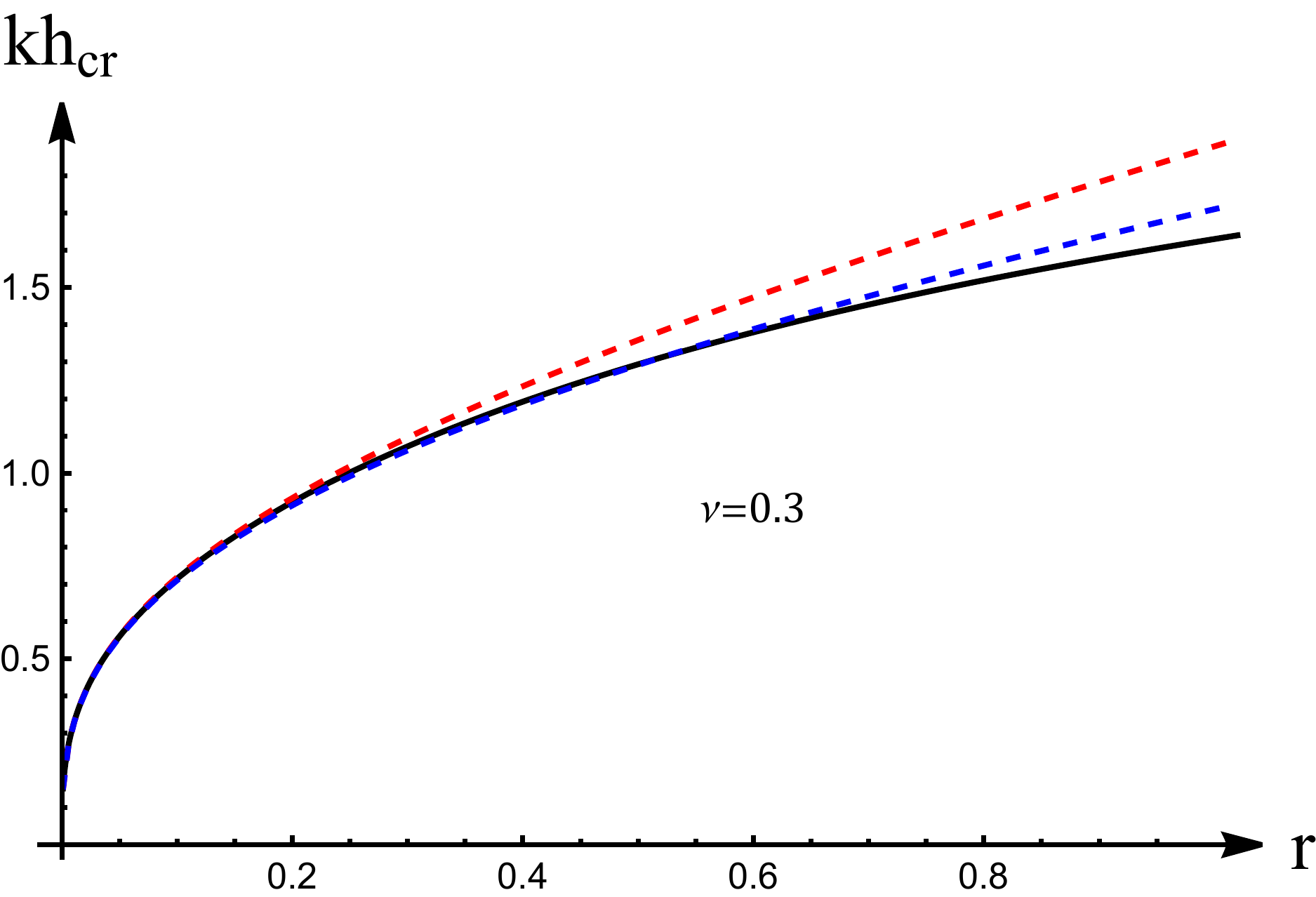} \\
(a) & & (b)
\end{tabular}
\caption{Comparison of asymptotic results (dashed lines) with exact results (black solid line) when $\nu=0.3$. The red dashed lines in (a) and (b) correspond to \rr{app4} and \rr{kh}, respectively, and whereas the blue lines represent lower order approximations with $g_3$ and $g_5$ set to zero.}
\label{h1}
\end{center}
\end{figure}

  \begin{figure}[ht]
\begin{center}
\begin{tabular}{ccc}
\includegraphics[scale=0.35]{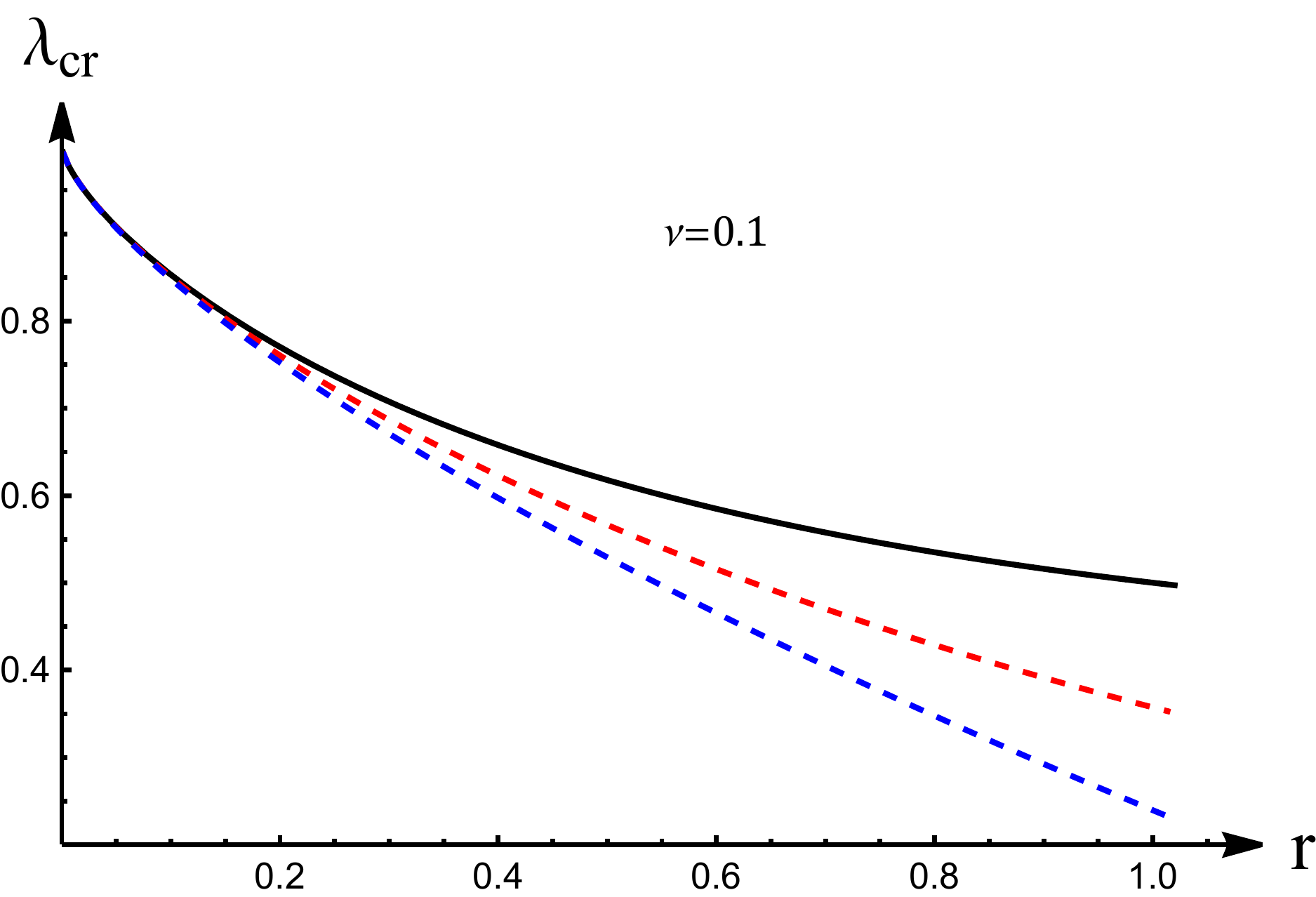}& & \includegraphics[scale=0.35]{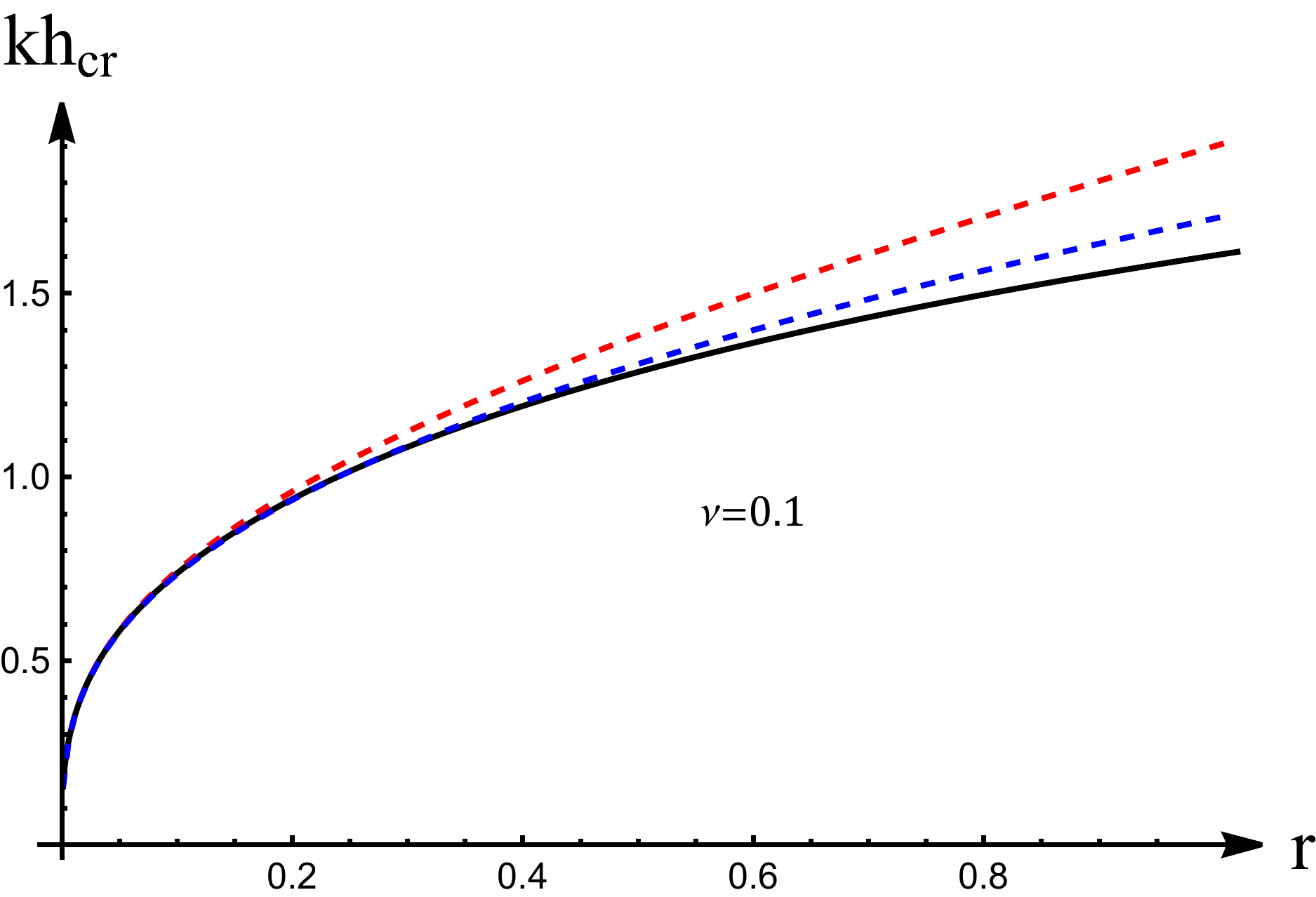} \\
(a) & & (b)
\end{tabular}
\caption{Comparison of asymptotic results (dashed lines) with exact results (black solid line) when $\nu=0.1$. The red dashed lines in (a) and (b) correspond to \rr{app4} and \rr{kh}, respectively, and whereas the blue lines represent lower order approximations with $g_3$ and $g_5$ set to zero.}
\label{h1}
\end{center}
\end{figure}

\section{Conclusion}
Reduced plate models have played an important role in engineering applications since they allow the most important information to be extracted without having to solve the fully three dimensional elasticity problem. In particular, the Euler-Bernoulli beam theory has frequently been used in studying pattern formation in film-substrate bilayers. Together with the Winkler assumption for the response of the substrate, this classical model is self-consistent and yields a leading order asymptotic expansion for the critical strain that agrees with that given by the exact 3D theory. In this paper, we have derived a refined model for the film-substrate interaction. The main motivation for deriving such a model is our planned study of pattern formation in a coated half-space where the coating has periodic and piecewise homogeneous material properties. For this problem, it is much harder to use the exact 3D theory to obtain analytical results than for the much studied case when the coating is homogenous. Our preliminary study has shown that using the classical model is an attractive option, but its effectiveness is seriously hampered by the limited range of validity in which the theory is valid. It is hoped that the current refined model will increase the range of validity and the relevant results will be published in a sequel to the current paper.

We used the methodology first proposed by \cite{ds2014} to derive the relationship between the traction and displacement vectors at the interface. The derivation consists of two steps. The first step is to express the traction and displacement vectors at the interface, ${\bm t}(x_1, h)$ and ${\bm u}(x_1, h)$,  in terms of the displacement vector at the traction-free surface, ${\bm u}(x_1, 0)$. The second step is to treat the expression for ${\bm u}(x_1, h)$ as an asymptotic expansion in $h$ and invert it. The most attractive feature of this methodology is that all manipulations only involve the solutions of linear algebraic equations and the expansions can be carried out to any order in $h$ on a symbolic manipulation platform. Also, our treatment of both the coating and half-space can be easily generalised to deal with anisotropic and/or inhomogeneous materials \citep{spencer1984, spencer2005}, nonlinear effects \citep{erbay1997}, liquid crystal elastomers \citep{gm2021,lmd2021}, and the case when $U$ and $V$ depend on both of the in-plane coordinates.


\subsection*{Acknowledgements}
We thank Professors Julius Kaplunov and John Chapman at Keele University for useful discussions.
This work was supported by the National Natural Science Foundation of China (Grant Nos 12072224 and 12072227).


\end{document}